\documentclass[12pt,preprint]{aastex}

\newcommand{\vsini}{\mbox{$v~\sin i$}}

\newcommand{\vinf}{\mbox{$v_{\infty}$}}

\newcommand{\Halpha}{\mbox{H$\alpha$}}

\newcommand{\kmsec}{\mbox{$\rm{km \; s^{-1}}$}}

\newcommand{\EM}{\mbox{$EM$}}

\newcommand{\mdot}{\mbox{$\dot{M}$}}

\newcommand{\Rstar}{\mbox{$R_*$}}

\newcommand{\etal}{\mbox{et~al.~}}

\newcommand{\Berghofer}{\mbox{Bergh\"{o}fer~}}

\shorttitle{X-ray Emission from MTD of Oe/Be stars}

\shortauthors{Li \etal.}

\begin{document}

\title{X-ray Emission from Magnetically Torqued Disks of Oe/Be Stars}


\author {
Q. Li\altaffilmark{1,3}, 
J.~P. Cassinelli\altaffilmark{2}, 
J.~C. Brown\altaffilmark{3}, 
W.~L. Waldron\altaffilmark{4}, 
N.A. Miller\altaffilmark{5}
}

\altaffiltext{1}{Dept. of Astronomy, Beijing Normal University, Beijing 100875, China; qkli@bnu.edu.cn}
\altaffiltext{2}{Dept. of Astronomy, University of Wisconsin-Madison, 53711; cassinelli@astro.wisc.edu}
\altaffiltext{3}{Dept. of Physics and Astronomy, University of Glasgow, Glasgow, G12 8QQ, Scotland, UK; john@astro.gla.ac.uk, li@astro.gla.ac.uk}
\altaffiltext{4}{Eureka Scientific, Inc., 2452 Delmer Street Suite 100, Oakland, CA 94602-3017; wwaldron@satx.rr.com}
\altaffiltext{5}{Dept. of Physics and Astronomy, University of Wisconsin-Eau Claire, Eau Claire, WI, 54702; millerna@uwec.edu}

\begin{abstract}
The near Main Sequence B stars show a sharp drop-off in their X-ray to
bolometric luminosity ratio in going from B1 to later spectral types. Here
we focus attention on the subset of these stars which are also Oe/Be stars,
to test the concept that the disks of these stars form by magnetic channeling 
of wind material toward the equator. Calculations are made of the X-rays expected 
from the Magnetically Torqued Disk (MTD) model for Be stars discussed by Cassinelli \etal (2002), 
by Maheswaran (2003), and by Brown \etal (2004). In this model, the wind outflow from 
Be stars is channeled and torqued by a magnetic field
such that the flows from the upper and lower hemispheres of the star collide
as they approach the equatorial zone. X-rays are produced by the material that enters the 
shocks above and below the disk region and radiatively cools and compresses 
in moving toward the MTD central plane. It differs from the Babel \& Montemerle (1997) 
model in having a weaker B field and a large centrifugal effect. The dominant 
parameters in the model are the $\beta$ value of the velocity law, the rotation rate 
of the star, $S_o$, and the ratio of the magnetic field energy density to the disk 
gravitational energy density, $\gamma$.

The model predictions are compared with the $ROSAT$ observations obtained
for an O9.5 star $\zeta$ Oph from \Berghofer\ \etal (1996) and for 7 Be stars
from Cohen \etal (1997). Two types of fitting models were used to compare
predictions with observations of X-ray luminosities versus spectral types.
In the first model we choose an estimate of the spin rate parameter $S_o$
from observed $\vsini$ values, and choose $\gamma$ using the threshold
magnetic field value derived in Cassinelli \etal (2002), then the $\beta$
value is adjusted to fit the observations. For all but one case, the
$\beta$ value was found somewhat larger than unity, a typical value derived 
for radially streaming stellar winds. This value, appropriate to a slowly 
accelerating wind, might be an indication that the magnetic field modifies the dynamics of the 
outflow from the star. In the second fitting model, we choose $\beta$ to be 
unity, $\gamma$ as in the first model, and adjust $S_o$ to fit the observations. 
For these comparisons with
the X-ray observations we find that $S_o$ is in the range 0.49 to 0.88,
which agrees with traditional estimates of the rotation rate of Be stars,
but is below the breakup values that are required in recent non-magnetic
models for Be star disks (Townsend \etal 2004). 

Extra considerations are also given here to the well studied Oe star $\zeta$ Oph 
for which we have $Chandra$ 
observations of the X-ray line profiles of the triad of He-like lines from the ion Mg XI. 
We find that a reasonably good fit is made to the observed Mg XI line profiles. The lines are predicted to 
form primarily at radial distance of about two stellar radii in the disk, and the ratio of the forbidden to intercombination lines (i.e., $f/i$) that is a
diagnostic of source distances, agrees with this prediction. In addition, the
lines are broad, with HWHM of about 400 $\kmsec$. Again this is in 
compatibility with the model predictions for the disk rotation of this star. 

Thus the X-ray properties add to the list of observables which can be
explained within the context of the MTD concept. This list already
includes the \Halpha\ equivalent widths and white light polarization of Be
stars. We do not include here the disk density correction due to gravity, 
neglected in Cassinelli \etal (2002) but included in Maheswaran (2003).
This will likely increase the \Halpha\ and
polarization predictions but have little or no effect on the X-rays
which are generated in the upstream region. Thus the X-rays are not sensitive
to the cooled disk region where Keplerization of the disk material might
be occurring. Nonetheless the process by which matter and angular momentum
are added to the disk are equally important and this 
study indicates that X-ray properties are consistent with the overall MTD concept.

\end{abstract}


\section{Introduction}

In a survey of the X-ray emission of near Main-Sequence B stars (or B
V stars), Cassinelli \etal (1994) and Cohen \etal (1997) found a
departure from the canonical ``law" relating X-ray luminosity to the
bolometric luminosity for hot star X-rays: $L_x/L_B=10^{-7}$. This
relation holds for stars throughout the O spectral range and extends
to about B1 V. However, beyond that there is a sharp drop in the ratio values 
in going to B3 V by about 2 orders of magnitude. Cohen \etal (1997) investigated
if the sharp decrease could be explained merely by the reduction of
the wind outflow from these stars, and found that this could
indeed explain the initial decrease in the X-ray luminosity, but that
in going to even later B V stars another problem arose. The emission
measures of the X-ray producing material at spectral type B3 V and
later become larger than the predicted wind emission measures for B
stars for the smooth wind case. Cohen \etal (1997) suggested
that this could be the first indication that the late B stars lie at
the transition to the outer atmospheric structure of cool stars, for
which surface magnetic fields control the X-ray properties. These
ideas were based on a spherical radial outflow picture for B stars.

In fact, B stars are known to be rather rapid rotators. Bjorkman \&
Cassinelli (1993) developed the Wind Compressed Disk (WCD) model, which
suggested the wind from a rapidly rotating star would orbit towards the
equatorial region where it would shock and compress the incident gas. The
model had success in explaining the polarization properties of emission line
Be stars (Wood \etal 1997). WCD was also supported, initially, by
hydrodynamic simulations performed by Owocki, Cranmer \& Blondin (1994).
However, in a more detailed consideration of the flow to the equator idea,
Owocki, Cranmer \& Gayley (1996) found that non-radial line forces in a
rotating and distorted star tend to impede the flow to the equator and
produce a bipolar flow instead. Hence, the cause of the {\it disks} that
exist around Be stars as opposed to {\it polar plumes}, has become a topic
of much debate among theorists.

Observers also found problems with the WCD idea. Hanuschik \etal
(1996) found in their observations of {\it equator-on} Be stars, that
the mass outflow speeds {\it detectable} in the disks were negligible
compared with the steady but slow equatorial outflows predicted by the
WCD model, though no one seems ever to have predicted whether in fact the WCD material
is dense enough to be detectable in this way. Even more interesting was their observational
conclusion that the azimuthal speed of the inner disk material was larger than the angular 
speed of the star from which the disk presumably originated (as it must be to remain 
in Keplerian orbit unless supported by other forces.) 
Specifically, observations of the equator-on Be star $\beta^1$ Mon showed that the Fe II line 
emission arising from the
disk is broader than the $\vsini$ value derived from photospheric
lines.  Hanuschik \etal\ (1996) suggested that the equatorial disk material
was in Keplerian motion about the star.  In the context of any Keplerian paradigm, 
it is especially important to note that in order to form a Keplerian disk, there needs 
to be an increase in the specific angular momentum of the matter after it leaves the star.  
The mechanism for providing that additional angular momentum is currently an unresolved 
subject of debate (e.g. Baade \& Ud-Doula 2005, and Brown \& Cassinelli 2005).

Transfer of mass and torquing of the outflow could be produced by
magnetic fields rooted in the star's surface, as is well-known from
magnetic rotator theory (Lamers \& Cassinelli 1999 Ch. 9).  The
existence of magnetic fields in hot stars is now well established
(Donati \etal 2001, 2002). Thus, the Magnetically Torqued Disk (MTD) model was proposed 
by Cassinelli et al. (2002) for the disks around
Be stars. In this model, the star is pictured as having a co-aligned
dipolar field that both channels and torques the wind from the star
towards a disk.  Being that the detailed structure of the magnetic
fields of these stars has not yet been determined, it seems reasonable
to use the pure dipole as the most conservative hypothesis. Minimal
magnetic fields were derived from the need to torque the outflow and
the much denser disk material to Keplerian speeds, and the fields
required were compared with upper and lower limits for hot star fields
that had been derived by Maheswaran \& Cassinelli (1988, 1992). For
stars of spectral class B2 V (which correspond to the most common
class of Be stars), the field required to torque the dense disk is
about 300 Gauss while that required to torque the wind is only around
10 Gauss. This difference is because the field needed to torque a wind
is proportional to the square root of the density, and the density of
the wind is about 3 orders of magnitude lower than the density at the
equatorial plane in the disk. The fact that the higher figure is
comparable to the fields that have now been derived from multi-line
Zeeman effect measurements of the very slowly rotating star $\beta$
Cephei (Donati \etal 2001) shows that both the flow and the disk will
in reality undergo MTD torquing  for this star.  In other stars,
the magnetic fields (whose strengths have not yet been measured) may
lie in an intermediate regime in which the wind material would be
torqued to high specific angular momentum while being channeled, but
have entirely Keplerian dynamics in the denser disk (Owocki \etal 2005). 
The MTD model was shown to produce the \Halpha\ emission
observed in Be stars (Doazan \etal 1991), and also the level of
intrinsic polarization seen (Quirrenbach \etal 1998). It 
is important to note that only a fraction of B stars are Be stars, and those 
stars identified as Be stars only spend a fraction of their time in a state with 
identifiable Be-star features. Therefore it is not necessary for a theoretical 
paradigm to cause a magnetically torqued disk for all possible sets of stellar and 
magnetic parameters, it is only necessary for a disk model to
encompass a wide enough range of parameter space so as to make it
reasonable that some stars show disks some of the time.  Detailed
comparison with observations will only be possible when the ``duty
cycles'' of Be stars and the Be star fraction have been better
determined observationally.

In the original MTD model, the stellar wind mass flux and wind speed
distribution were taken to be uniform over the stellar surface. However, in
the case of a rapidly rotating star, this assumption is invalid since the
rotation results in gravity darkening (Von Zeipel 1924) and reduces the wind
mass flux and the terminal speed in the equatorial region (Owocki \etal
1998). By incorporating Gravity Darkening into the MTD model (MTDGD), Brown \etal (2004)
derived several important disk properties such as, the dependence of the disk mass
density distribution on its extent, the total number of disk
particles, and the functional dependencies of the emission measure and
polarization on the rotation rate ($S_o$) and wind velocity law ($\beta$).
In contrast to what had been expected, they found that the critical rotation
(or $S_o= 1$) is not optimal for creation of hot star disks.

One important omission in the basic MTD formulation, as well as in MTDGD to
date was the erroneous neglect of gravity in the disk density
structure - see Brown \& Cassinelli (2005). Though $g_z \propto z$ is
zero in the equatorial plane, its increase with $z$ enhances the
density in the disk and causes it to grow with time. It will likely increase 
the \Halpha\ and polarization predictions. In turn, this will
result eventually in radial escape of disk material, though whether
this is slow and steady or episodic as claimed by Owocki \& ud-Doula
(2003) is not yet clear.

Other computational and analytic attempts to model disks similar
to those envisioned here have met with mixed results. Owocki \& ud-Doula (2003) 
criticized the MTD idea because the model
was found to be unstable in their MHD simulations.  However, the
numerical simulations by Keppens \& Goedbloed (1999, 2000), and Matt
\etal (2000) demonstrated the existence of disks around some hot stars
like post-AGB stars. Also Maheswaran (2003) has studied this scenario using analytic 
MHD and finds results that the magnetically torqued disks are likely to be persistent, 
which cast doubt on the numerical simulations of Owocki \& ud-Doula (2003).  Thus
further work is needed to test the basic ideas of the MTD and MTDGD
models, either thorough numerical and analytic MHD calculations, or using
observational diagnostics from radio to X-ray wavelengths.

In this paper, however, we are primarily concerned with
X-ray emission from the MTDGD models, emission which occurs
well upstream of the dense disk, and explain the X-ray anomalies associated
with Be stars. This will be little affected by the
density in the disk itself though our use of MTD to find the outer
disk radius will make our X-ray source emission measure estimates a
little too high. A successful model should be able to explain the
drop-off at B2 Ve, the apparently excessive X-rays of late BV stars, while using 
disk parameters consistent with theory 
and the entire set of observational data. This process can then be inverted to use X-ray properties of a star to derive limits on its wind, disk, and
rotational properties.  In Section 2, we describe how X-ray emission
is produced by the model. The effects of model parameters on X-ray
emission are discussed in Section 3.  Comparisons of model predictions
with both $ROSAT$ and $Chandra$ observations are presented in Section 4. 
The discussion and conclusions are presented in Section 5.

\section{Model for X-ray emission}

The basic concept for X-ray production by MTDGD
models is that there are shock heated regions above and below an
equatorial disk
where the winds from the upper and lower hemispheres of the star
collide. To establish the X-ray emission properties we need to
make model predictions of the density and temperature
structure in the post-shock regions. The disk and pre-disk densities
are dependent
on the mass and momentum flux from the corresponding latitude zones
of the stellar surface which are funneled
via magnetic flow tubes to the disk. The temperature depends on the speed at
which the matter collides with the shocks at the disk boundaries. We treat
these aspects in turn, and then discuss how we combine various parts
of the heated disk to determine the resultant X-ray spectrum and total X-ray
luminosity.

As in the previous MTD papers the dimensionless rotation rate
of the star $S_o$ is defined as the Keplerian fraction by
\begin {equation}
S_o=\sqrt{\Omega_o^2 R^3/GM},
\end {equation}
for stellar angular velocity $\Omega_o$. This determines the inner
and outer boundary locations of the disk and 
the effects of rotational Gravity Darkening on the disk. The ratio $\gamma$, 
which, together with $S_o$ determines the effect of the magnetic field 
on the disk and wind, was defined as
\begin {equation}
\gamma=\left(\frac {B_o^2/8\pi}{GM\rho_o/2R} \right)^{1/2}
\label{gamma}
\end {equation}
where $\rho_o=[\dot M/(4\pi R^2
v_{\infty})](v_{\infty}/c_s)^2$ is the characteristic density of the cool gas at the equatorial plane and $\mdot/(4\pi R^2 \vinf)$ is the characteristic density for the wind. Therefore, 
$\gamma$ is a measure of the magnetic field energy density relative
to the gravitational energy density of the wind material near the star. So 
a unit value for $\gamma$ provides an indication of the minimal field
needed to form a disk. The field then determines the latitude
range of the stellar flow that forms a disk, and the inner and outer radii.

The mass flux from the base of the wind is given in MTDGD theory by
\begin {equation}
F_m(x)=\frac {\mdot}{4\pi R^2 (1- \frac{2}{3} S_o^2)} x^{-3}
\{1-S_o^2[1-(1-\frac {1}{x})^2]\},
\end {equation}
and the wind speed
\begin {equation}
v_w (x)=\vinf \left(1-\frac{1}{x} \right)^{\beta} \left( 1-S_o^2
\left[ 1-(1-\frac {1}{x})^2 \right] \right) ^{1/2},
\end {equation}
where $x$ is the radial distance in the disk from the center of the star in units of
the stellar radius (i.e. $r/R_{\star}$), $\vinf$ is the terminal velocity of
the wind flow if it were unimpeded to infinity. Thus from $F_m$, the
pre-shock mass density
approaching the disk is
\begin {equation}
\rho_w=\frac{F_m}{v_w}=\frac {\mdot}{4\pi R^2 v_{\infty} } x^{-3}
\left( 1-\frac {1}{x} \right) ^{-\beta}
\frac{ \left( 1-S_o^2 \left[ 1-(1-\frac{1}{x})^2 \right] \right) ^{1/2}}
{1- \frac{2}{3} S_o^2}.
\end{equation}

We assume the disk is formed by the shock-compression above and below the
equatorial plane. Note, for simplicity, here strong, normal shocks are assumed, while the time-variable structures in the disks and the radiative overstability in the shocks (e.g., Pittard \etal 2005) are not taken into account in the model.
Resultant shock temperatures greater
than $10^6$ K will produce X-ray emission.

In terms of the jump
conditions, at the top of the disk the shock density is four times the wind
density, namely
\begin{equation}
\rho_s=4 \rho_w.
\label {rho}
\end{equation}

The temperature at the wind interface of the shocked disk is given by
\begin {equation}
T_s(x)=1.44\times 10^{7} ~ v_8^2(x),
\label {eqT}
\end{equation}
where $T_s$ is the shock temperature in degrees K and $v_8$ is the
incident wind speed in $10^8$ cm s$^{-1}$.

According to the standard X-ray models by Hillier \etal (1993) and later by
Feldmeier \etal (1997), the energy emitted per second per
Hz from a volume $dV$ in all directions is given by
\begin{equation}
d\epsilon_{\nu}= n_p n_e \hat {\Lambda}_{\nu} \left(T_s \right)dV,
\label {epsilon}
\end{equation}
where $n_p$ is the proton density, $n_e$ the electron density,
$T_s$ is the temperature reached at the wind-disk
shock located at $(r,\phi,z)$, using cylindrical geometry where $r$ is the 
radial distance in the equatorial plane, $\phi$ is the azimuthal angle and
$z$ is the distance above the equatorial plane. $\hat {\Lambda}_{\nu}$ in the
above equation (eq.~\ref{epsilon}) is determined by averaging across the cooling length as given by Feldmeier \etal (1997) as 
\begin{equation}
\hat {\Lambda}_{\nu} \left( T_s \right)
=\frac{1}{l_c} \int_z^{z-l_c} {\hat f}^2(z')
\Lambda_{\nu}\left(T_s(z'){\hat g} (z')\right) dz',
\end{equation}
where $\Lambda_{\nu}$ is the frequency dependent cooling function of a hot
plasma, $z$ is the location of the shock front, and $z'$ is the coordinate
in the cooling layer of extent $l_c$ which is related to the velocity and
density of post-shock gas, and the chemical composition. The functions $\hat f$
and $\hat g$ describe the normalized density and temperature stratification in
the post-shock region, respectively. Note, we neglect the plasma motions 
in the post-shock flow which can generate small-scale magnetic structure that may provide some magnetic support and thus reduce the post-shock compression.

\begin{figure}
\plotone{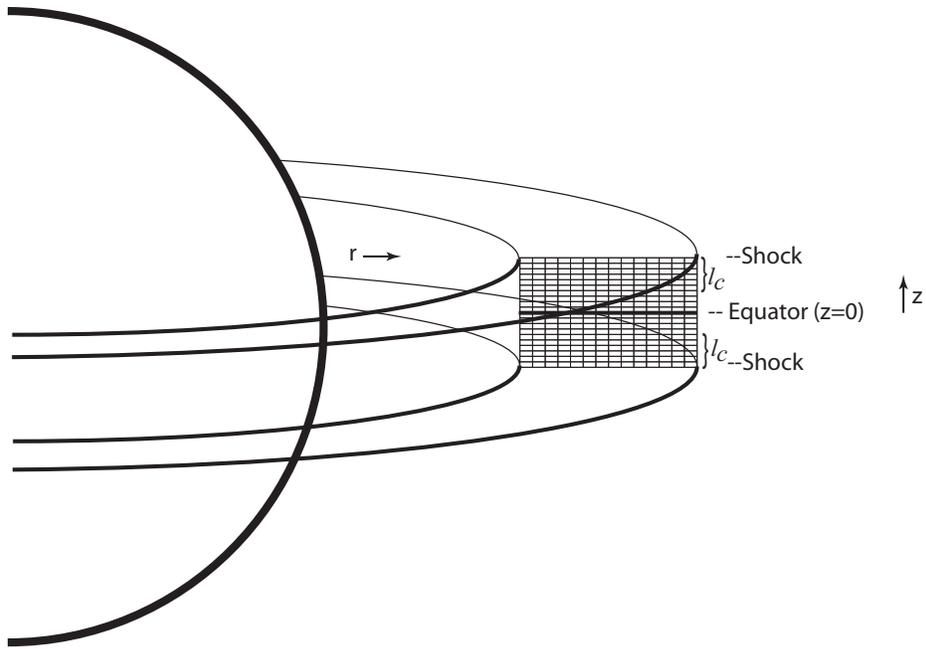}
\caption{A diagram which indicates the geometry of the shocked-disk
model (not-to-scale).
\label{f1}}
\end{figure}

In our treatment, we use the functional forms of $\hat f$ and $\hat g$ for
the temperature and density stratifications, respectively, as
defined in Feldmeier \etal (1997) who considered plane parallel shock fronts.
However, here the cooling layer above disk of Be stars
is split into many concentric rings (e.g. $\sim$ 200 rings) along the disk radial 
extent and each ring is sliced into vertical sub-layers (e.g. $\sim$100 sub-layers), 
as is described in Fig.~\ref{f1}. Thus, each sub-layer has a 
specific density and temperature, which are assumed constant throughout
the sub-layer.  With these, one may obtain the emission measure for each
ring and sub-layer from

\begin{equation}
\Delta (\EM)_i=n_e n_p \Delta V_i=\frac{\rho^2_i}{m^2_H\mu_e\mu_p} \Delta V_i,
\end{equation}
where $\Delta V_i$ and $\rho_i$ are the volume and mass density of the $i$th
sub-layer, respectively. $\mu_e$ and $\mu_p$ are mean particle weights per
electron and proton, respectively. Then, the X-ray emission from
each ring and sub-layer
is given (with $\Delta V_i = 2\pi r_i \Delta z_i \Delta r_i$)  by
\begin{equation}
\Delta L_{\nu,i}=\Delta (\EM)_i ~\Lambda_{\nu}(T_i),
\end{equation}
where $\Lambda_{\nu}(T_i)$ is the cooling function at temperature $T$,
frequency $\nu$ and chemical abundance (assumed solar), as
given by Astrophysical Plasma Emission Database (APED) described by
Smith \etal (1998).

The total X-ray emission for the entire shocked disk is found by
summing the emission over all the rings and sub-layers and by
multiplying by two
to account for the upper and lower shock fronts of the disk,
\begin{equation}
L_{\nu}=2\sum_i \Delta L_{\nu,i}.
\end{equation}

Integrating over the frequency range concerned, we obtain the X-ray
luminosity 
\begin{equation}
L_x=\int_{\nu_1}^{\nu_2} L_{\nu} d \nu,
\end{equation}
where $\nu_1$ and $\nu_2$ are the lower and upper X-ray frequencies 
in the energy band of the instrument, such as 0.1 keV to 2.4 keV 
in the case of $ROSAT$, for example. 

Calculations were carried out with sufficient numbers of rings
and sub-layers so that the results were no longer dependent on the specific
number of rings and sub-layers.

\section {Effects of Model Parameters on X-ray Emission}

The MTDGD models have been carried out for main sequence stars with
given effective temperatures, mass loss rates, and terminal
velocities.  Our interest is in the X-ray properties at each spectral
class as they are affected by the unknown wind velocity profiles, the
stellar rotation rate and the magnetic fields.  Thus, there are
essentially three free parameters for any given spectral type, the $\beta$ value of the velocity
law, the rotation rate parameter $S_o$, and the value of $\gamma$. For
other properties that can affect the X-ray emission, such as the mass
loss rate and terminal velocity, we choose values typical for each
spectral type as in previous papers dealing with the MTD model.  

\begin{figure}
\plotone{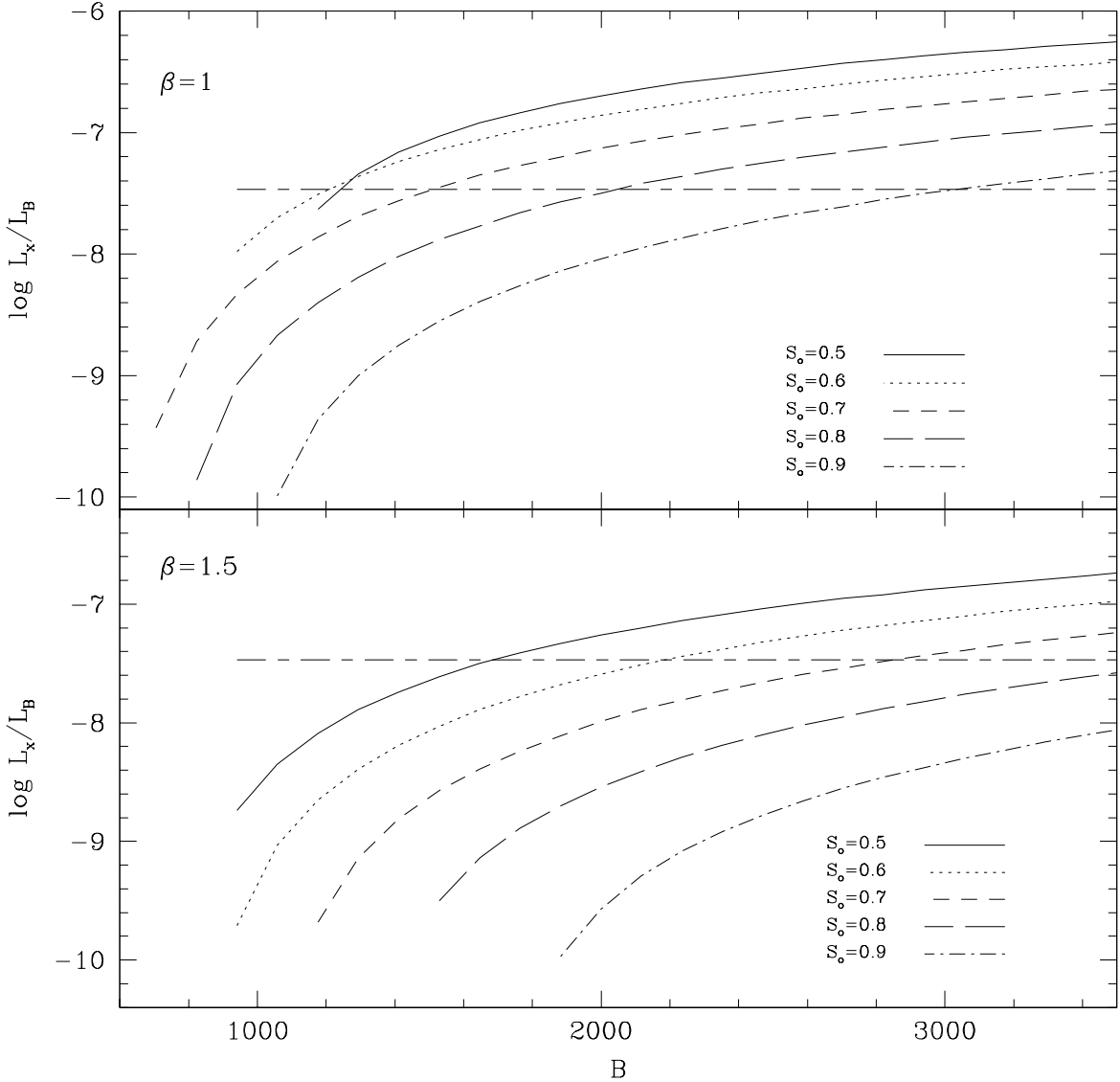}
\caption{The upper panel of the figure is the MTDGD model results for
log $L_x/L_B$ versus magnetic field B (i.e. $\gamma$) in Gauss for $\zeta$
Oph stellar parameters and for $\beta=1$ with the various $S_o$ indicated in
the figure. The curves show that log $L_x/L_B$ increases with increasing
field strength. This is because larger magnetic fields give rise to more
matter being torqued and channeled into the disk. Also note that as $S_o$
increases, log $L_x/L_B$ decreases and this is due to the gravity darkening
effects. Note, the horizontal line in the figure indicates the observational
value of log $L_x/L_B=-7.47$ for $\zeta$ Oph from $ROSAT$. The lower panel
of the figure is the same as that in the upper panel but now $\beta=1.5$
instead.
\label{f2}}
\end{figure}

\begin{figure}
\plotone{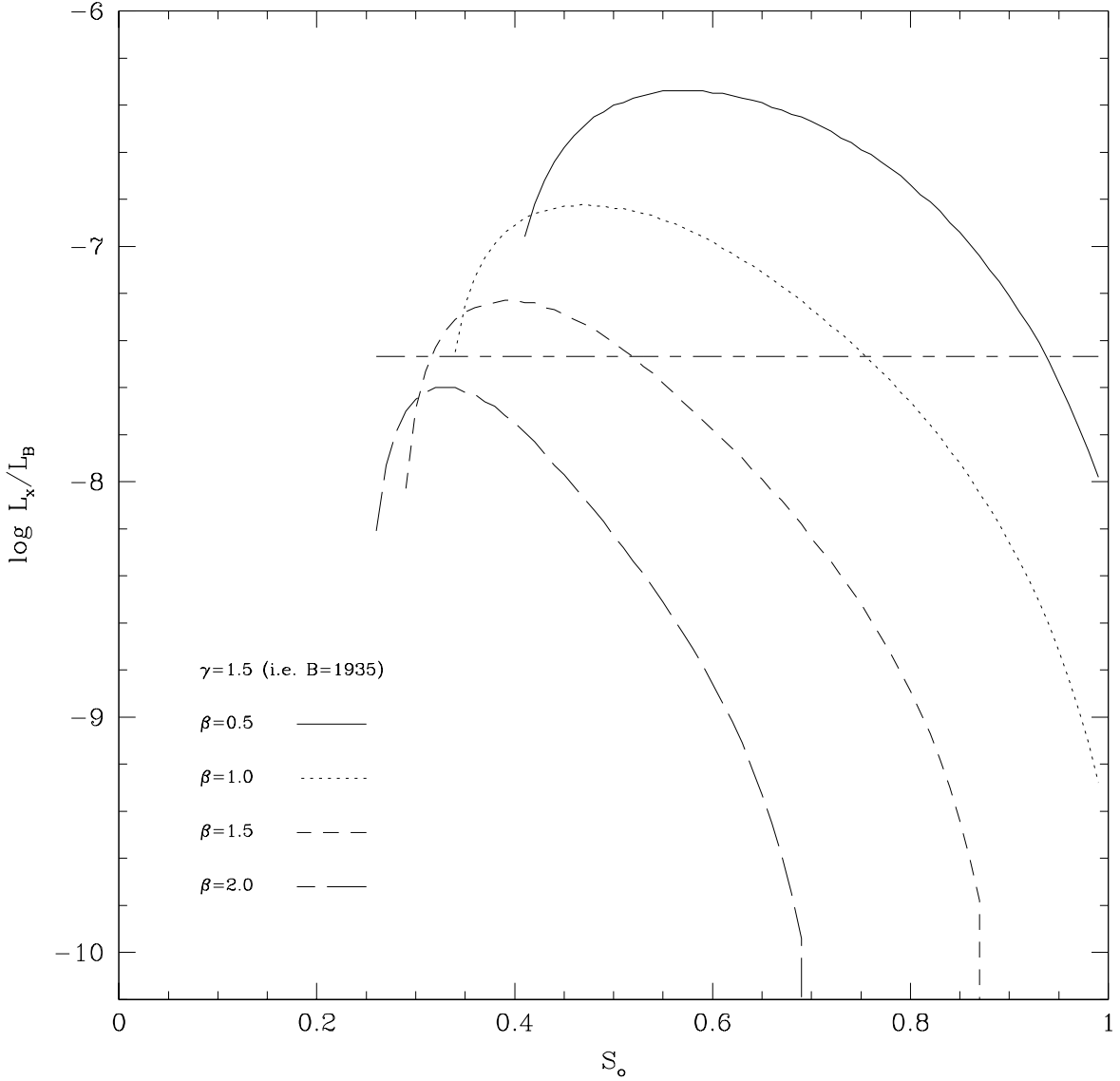}
\caption{The MTDGD model results of
Log $L_x/L_B$ versus $S_o$ for a given magnetic field $B=1935$ Gauss (which
corresponds to $\gamma=1.5$) as is needed for $\zeta$ Oph, and various
$\beta$ indicated in the figure. The curves show that as $S_o$ initially
increases from small values, the rotation helps to form the disk, and the
larger the rotation rate the greater the X-ray emission. However, when $S_o$
is larger than a turn-over value, the gravity darkening plays a dominant
role and leads to a decrease in the X-ray emission. Note also that $L_x/L_B$
tends to be larger for the smaller $\beta$ values, which correspond to
faster wind impacting the inner parts of the disk. The horizontal line
indicates the observational value of log $L_x/L_B=-7.47$ for $\zeta$ Oph as
obtained from $ROSAT$. 
\label{f3}}
\end{figure}

Each of the three model parameters (i.e., $\beta$, $S_o$, and $\gamma$) affects the X-ray emission
measure and $L_x/L_B$ ratio in different ways. To explain these various
effects, we discuss the results from our modeling of the star $\zeta$ Oph,
for which we find the following:

\begin{enumerate}

\item

Changing the velocity law $\beta$ value can significantly affect the
$L_x/L_B$ ratio. Changing $\beta$ has a small effect on the disk extent
but it significantly affects the X-ray source as a result of
the dependence of the cooling length on the wind velocity as parameterized
with $\beta$.  The cooling length depends on velocity as:
$l_c \propto v_w^4/\rho_s \propto v_w^5$ (see Feldmeier et~al. 1997).
Therefore, the overall effect on the emission measure is
$\EM \propto \rho_s^2 l_c \propto \rho^2_s v_w^4/\rho_s \propto v^3_w$.
If, for example, $\beta$ is increased, meaning that a more slowly accelerating
wind is incident upon the disk, then
the emission measure from shock regions becomes lower and in turn
$L_x/L_B$ is made lower, as is shown in Figs~\ref {f2} and \ref{f3} for
various $\beta$.

\item

Changing $\gamma$ also gives rise to a change in $L_x/L_B$. Increasing $\gamma$ leads to 
an increase in the radial extent of the disk, and the extension occurs
primarily outwardly away from the star. Thus there is a greater
radial range of emitting material channeled to the disk. In Fig.~\ref{f2},
we show the dependence of $L_x/L_B$ on the magnetic parameter $\gamma$.

\item

Increasing the rotation rate parameter $S_o$ affects the $L_x/L_B$ in three
ways: (a) while the rotation rate is smaller than a turn-over value (for example, 
for given values of $\beta=1$ and $\gamma=1.5$, this turn-over value is $S_o\sim 0.5$), 
the gravity darkening is negligible, so increasing rotation helps to form the disk. 
Hence, the greater the rotation rate the larger the amount of matter channeled to the disk; 
(b) as the rotation rate increases
further, the disk inner radius gets closer to the star, therefore a
slower and denser (so cooler) wind reaches the disk. Hence, we find a situation 
similar to that discussed above regarding the $\beta$ value, and $L_x/L_B$ actually decreases; 
(c) if $S_o$ is increased even further, beyond a turn-over value, the gravity darkening effect 
plays an important role in significantly reducing the mass flow to the disk from equatorial regions 
on the star, and  of course $L_x/L_B$ decreases significantly. Figs~\ref{f2} and 
\ref{f3} also show this trend for given $\gamma$ and $\beta$.

\end{enumerate}

In the tentative test, we show that when determining the $L_x/L_B$ from the model, we have to assume two of the three free parameters fixed to some values, then find how the $L_x/L_B$ varies with the remained one and whether the value of the remained one is appropriate for given the observed value of $L_x/L_B$. Figs.~\ref{f2} and \ref{f3} show the varying trends, and imply the probable values for these three free parameters. Strictly, the ranges of the parameters could only really be solved by
using a search to define the surface in 3-D ($S_o$, $\gamma$ and $\beta$) space with which fits are acceptable. Such a search will be carried out in the future and it may be more tightly constrained by fitting not only the X-ray luminosity but also the X-ray spectral hardness. From the above results, we might derive for our program stars like $\zeta$ Oph
a least square formula that would look like: $L_x = k * S_o^p\, \gamma^q\, \beta^w$, where constant $k$ 
is a certain fraction of the wind kinetic energy converted to X-rays and the precise amount depends 
on the spin rate $S_o$, B and $\beta$. By doing a numerical partial derivative for each power to fit 
our results, we would find the range of the three powers $p\sim 1-4$, $q\sim 5-8$, and $w\sim 2-5$, which 
shows $L_x$ seems to be super dependent on the value of $\gamma$. This may explain why we are 
able to fit all the stars so exactly by our model.

\section {Model Calculations and Comparisons with Observations}

The physical quantities of the 8 program stars are listed in
Table~\ref{tab1}.  We ran several models with considerations and compared the results with observations. 
The comparisons are shown as follows.

\subsection {Comparisons with $ROSAT$ observations}

For all program stars, we took MTDGD models to compute $L_x$
in three energy bands of $ROSAT$: soft (0.1--0.4 keV), hard (0.5--2.0
keV) and entire (0.1--2.4 keV) and got the hardness ratio
HR=(H-S)/(H+S), the emission measure, and the ratio of X-ray to
bolometric luminosity log $L_x/L_B$, which can be compared with the $ROSAT$
observations. In these calculations, we can choose various parameters
of $\beta$, $S_o$ and $\gamma$. But for simplicity, we always use the
minimal magnetic fields to $\gamma$ as given by Cassinelli
et~al. (2002), then  fix either
$\beta$ or $S_o$ and let the other one be adjustable. Note, in fact, what we did is 
to adjust parameters $S_o$ and $\beta$ until the fit is perfect. Hence we are assuming 
the model and inferring a range in $S_o$ and $\beta$ that is acceptable.

From the calculations with the MTDGD model, we achieve the observed
X-ray $L_x/L_B$ ratio, within the allowed range of adjustments of the
model's free parameters. In fitting the observed log $L_x/L_B$, we have
taken into account the following two different approaches. 

\begin{figure}
\plotone{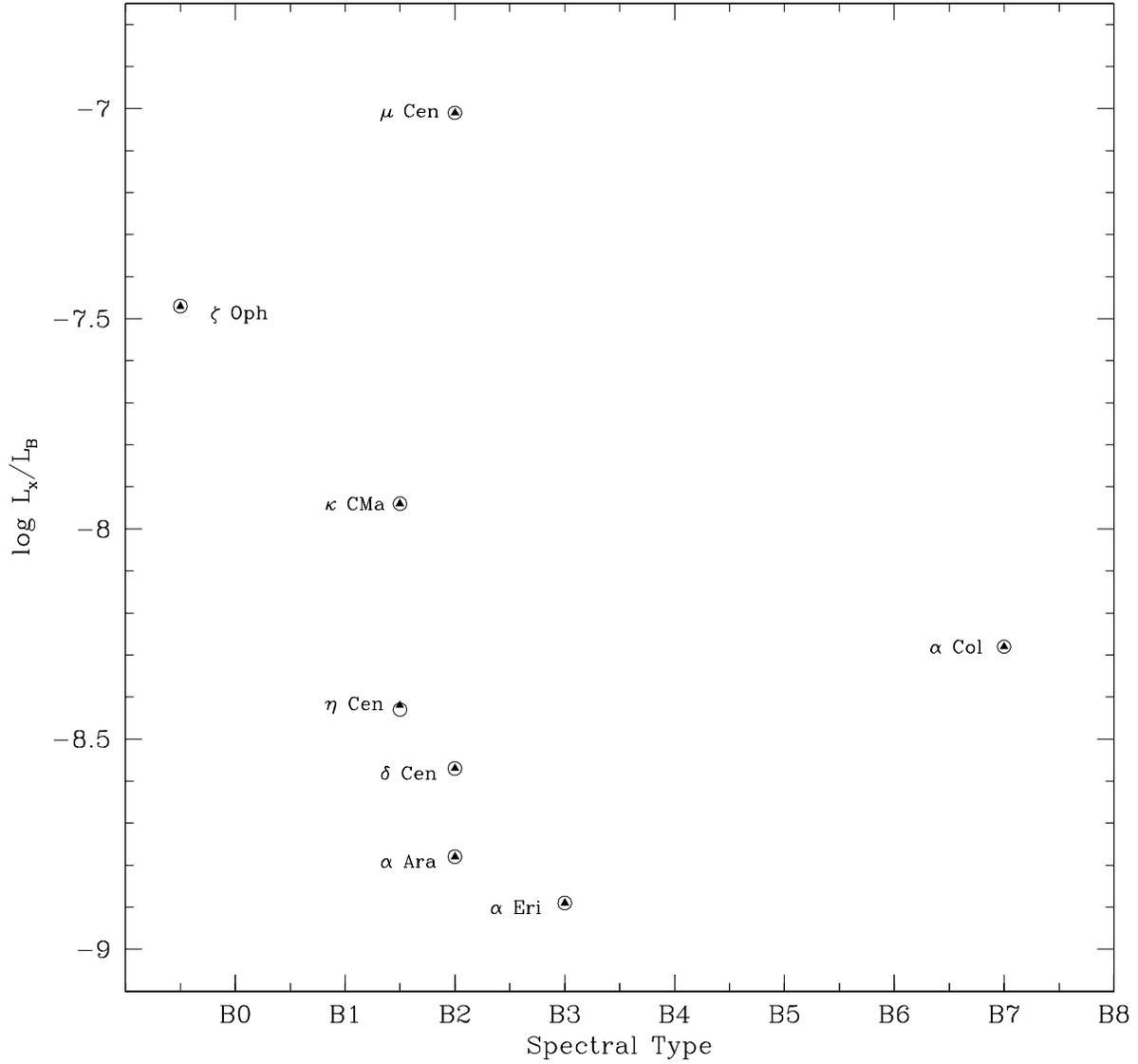}
\caption{Log $L_x/L_B$ versus spectral type. The model and
observational results with opened circle for the former and filled triangle
for the later, are plotted with the star names marked in the figure. The data
used is from Table \ref{tab2}. Note, we actually adjust 
parameters $S_o$, $\gamma$, and $\beta$  until the fit is perfect so that we
infer a range in $S_o$, $\gamma$, and $\beta$ that is acceptable.
\label{f4}}
\end{figure}

\begin{figure}
\plotone{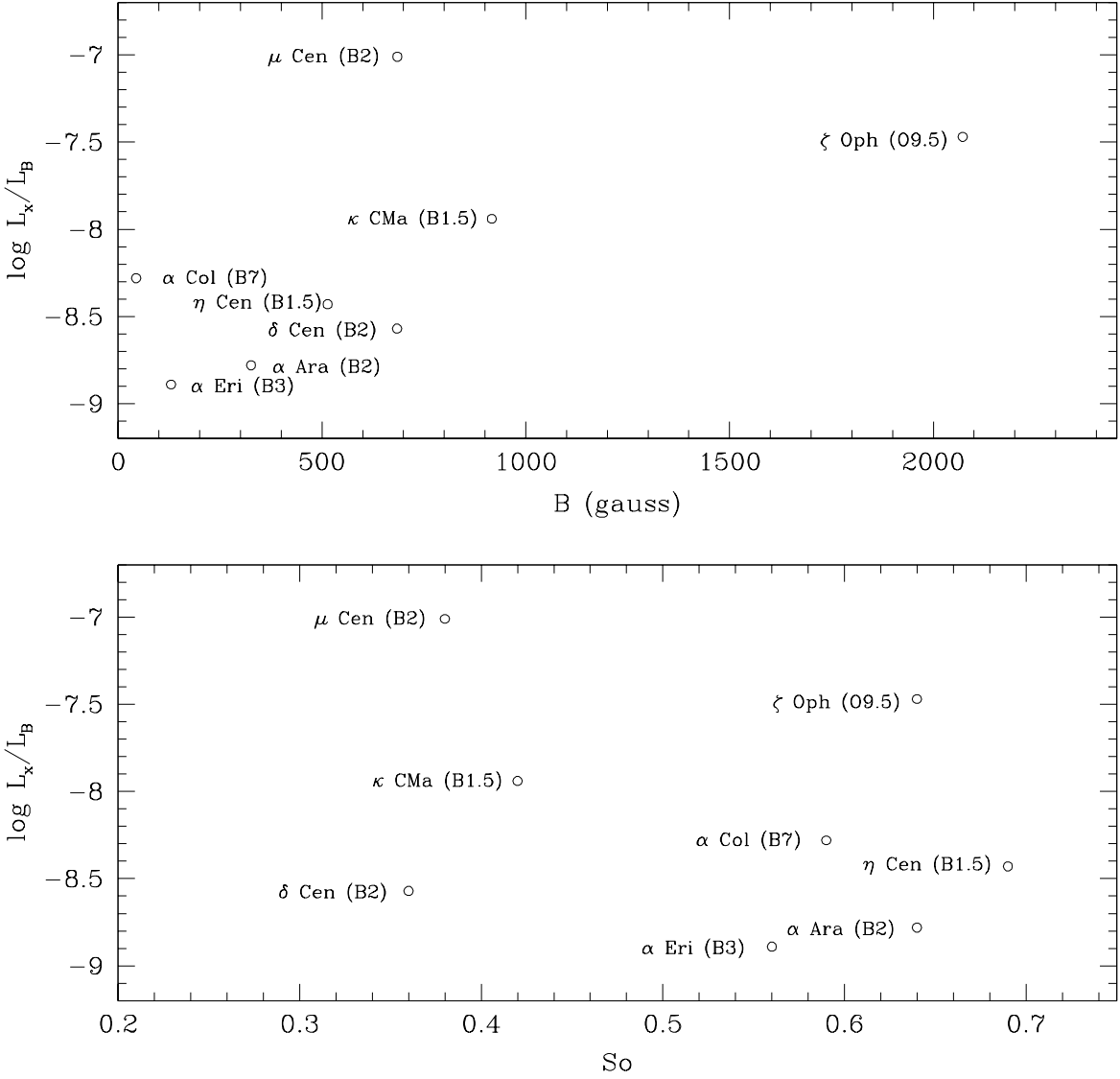}
\caption{The upper panel shows Log $L_x/L_B$ versus the magnetic fields
needed to produce from the MTDGD models discussed in this paper. In the
lower panel is shown the X-ray ratio versus the rotation rate parameter
$S_o$ that we find is needed. The star names are marked in the figures. The
B-star spectral type is included with each star's name in parentheses. The
data used is taken from Table \ref{tab2}. Note again we in fact adjust 
parameters $S_o$, $\gamma$, and $\beta$  in a range that is acceptable until the fit is perfect.
\label{f5}}
\end{figure}

In the first approach, we use the observed projected velocity $\vsini$ and used the average value of $<{\rm sin}~i>$ for
a random set of inclination angles to estimate a surface rotation speed.
This allows us to estimate the rotation rate $S_o$ of each star. This kind of
approach has been used by \citet{chandra} and later by many authors such as
\citet{port} to analyze the rotation of Be stars. Also in the first approach
to comparing the models with observations, we used the threshold (i.e.
minimal) magnetic field given by Cassinelli \etal (2002) as our tentative
field strength. Thus, for this case, we have one adjustable parameter ---
the velocity law index $\beta$ --- to provide a fit to the observations. As
is seen in Table~\ref{tab2}, with the exception of just one star, the value
of $\beta$ needed for the program stars is larger than unity. These values
correspond to slowly accelerating winds as compared with estimates of
$\beta$ for spherical winds. The calculation results are shown in Table~\ref{tab2} in
which we can see that the model results of the emission measure and
$L_x/L_B$ fit
the observations quite well for a given set of free parameters. We plot
figures with $L_x/L_B$ versus spectral type as in Fig.~\ref{f4} and
$L_x/L_B$ versus magnetic field B and $S_o$ as in Fig.~\ref{f5} in terms
of the model results in Table~\ref{tab2}. We also list the derived
hardness ratio of X-rays with regards to certain disk properties in the
table and find it marginally in agreement with observations. We see that the hardness 
ratio of $\zeta$ Oph ($HR_t=0.630$) fit the observation
($HR_o=0.908$) marginally well.  Unfortunately, the observational hardness
ratios for other program stars are not available from $ROSAT$, but we still
list the model results in the table for comparisons in the future.

As a second approach, we follow the radiative-driven wind theory and simply
assume $\beta=1$, then we can fit the observed $L_x/L_B$ by adjusting the
remaining parameter $S_o$. Note, the threshold field used here is determined
by $\beta$ and $S_o$ for a given star as in Cassinelli et~al. (2002). The
calculation results are shown in Table~\ref{tab3} in which we can find that
the range of $S_o$ for the various stars is from 0.49 to 0.88. This
range is consistent with traditional values of the rotation rates
for Be stars  \citep {under,slett}.

Thus we have found in using two approaches that the MTDGD
models provide good fits to the X-ray observations. If one
could find either a better way for estimating $S_o$ or
$\beta$ for these channeled winds, then it would in principle
be possible to derive new information on the process by which
co-rotating magnetic fields actually produce disks.
It is at least clear that the idea that angular
momentum is transferred to disks by magnetic fields is broadly
consistent, when Gravity Darkening is included, with the X-ray observations of
Oe/Be stars.

\subsection {Comparisons with $Chandra$ observations of $\zeta$ Oph}

The Oe star $\zeta$ Oph has been observed at high spectral resolution with the
$Chandra$ Satellite (Waldron 2005).  The interpretation of the X-ray emission line profiles using emission line
 ratio diagnostics has led to a significantly improved understanding of
 the nature of the X-ray sources. This is especially in comparison with
 the information we had from earlier studies using low spectral resolution
 satellites.  The observed line widths now provide information about the
 spread in line of sight velocities associated with the X-ray sources. The
 $f/i$ line ratios obtained from He-like ion $fir$ (forbidden, intercombination, resonance)  emission lines  provides a diagnostic of the radial locations of the shock source regions for these ions (e.g., Kahn \etal 2001; Waldron \&
Cassinelli 2001, 2007). Here we use the MTDGD model for $\zeta$ Oph to find the source region associated with the ion Mg XI and compare the line
 profile that would form in this region versus observations of the line
 from the $Chandra$ High Energy Transmission Grating Spectrometer (HETGS).

 The primary goal of this exercise is to demonstrate that the MTDGD
 concept can reproduce the general properties of the Mg XI $fir$ lines
 without adjusting fitting parameters. The MTDGD predicts that the Mg XI
 emitting region is an annular region above the disk at a determined radius
 where temperatures reach values of order 5 MK at which Mg XI 
 can form. We find that this is at about 1.8 $\Rstar$. The model that is used 
 to calculate the line profile assumes a ring of emission at 1.8 $\Rstar$
 and uses the post-shock density and temperature at that location. For comparison, the MTDGD radial location is about 22\% larger than the
source location obtained by Gagne et al. (2005) (1.2 to 1.4 $\Rstar$)
in their model of the young magnetic O-star $\theta^1$ Ori C.  From
 model predictions from Table~\ref{tab2}, we know the value of $S_o$ and thus
 the angular speed at the Mg XI formation region. From Table~\ref{tab1} we know
 the value of $\vsini$ of the surface. Thus we can use the observed line
 width, assumed to be from the orbital velocity of the source region
 to derive the inclination factor $\sin i$. This corresponds to an
angle $i = 53^{\circ}$.  The predicted X-ray source temperature of the ring
of emission is determined from equation~\ref {eqT} using the incident wind speed
determined by $\vinf$ from Table~\ref{tab1} and assuming $\beta = 1.29$. This yields
$\rm T = 4.94$ MK, which is near the temperature of 5 MK at which the
 Mg XI ion has its maximum ion fraction.

A key feature of our model line calculations is that we include ``real"
temperature and density dependent line emissivities which means that there
is only ``one" normalization applied to the total line complex calculation
(i.e., we do not apply individual line normalizations).  Our emissivities
are determined by the MEKAL plasma emission code (Mewe \etal 1995).  The
main advantage of this code is that it allows one to explore density
sensitive lines which is an important aspect in studies of the He-like fir
line formation process in early-type stars.  It is well established that the
observed behavior of the He-like $fir$ lines, in particular the relative
strengths of the $i$- and $f$-lines (i.e., the $f/i$ line ratio), is not due to
density effects, but instead is dependent on the strength of the EUV/UV
photospheric flux (e.g., Kahn \etal 2001; Waldron \& Cassinelli 2001, 2007).
Although current emissivity codes do not provide a means for determining
line emission dependencies on EUV/UV flux (as first demonstrated by
Blumenthal \etal 1972), one of us (Waldron) developed a special algorithm
to simulate the $f/i$ ratio dependence on EUV/UV.  Since we know how the $f/i$
ratio dependence separately on density and on EUV/UV flux, 
by equating these
two relationships we can determine, what we call, an ``effective EUV density"
for any radial wind location for a given input photospheric EUV/UV flux.
This effective EUV density is then used in the MEKAL code to determine the
relative strengths of the $i$- and $f$-lines.  We point out that the actual
value of this effective EUV density is ``not" an actual physical density, it
is only a parameter that is used to simulate the effects of the EUV/UV flux
on the $f/i$ ratio.

\begin{figure}
\vspace*{-13mm}
\plotone{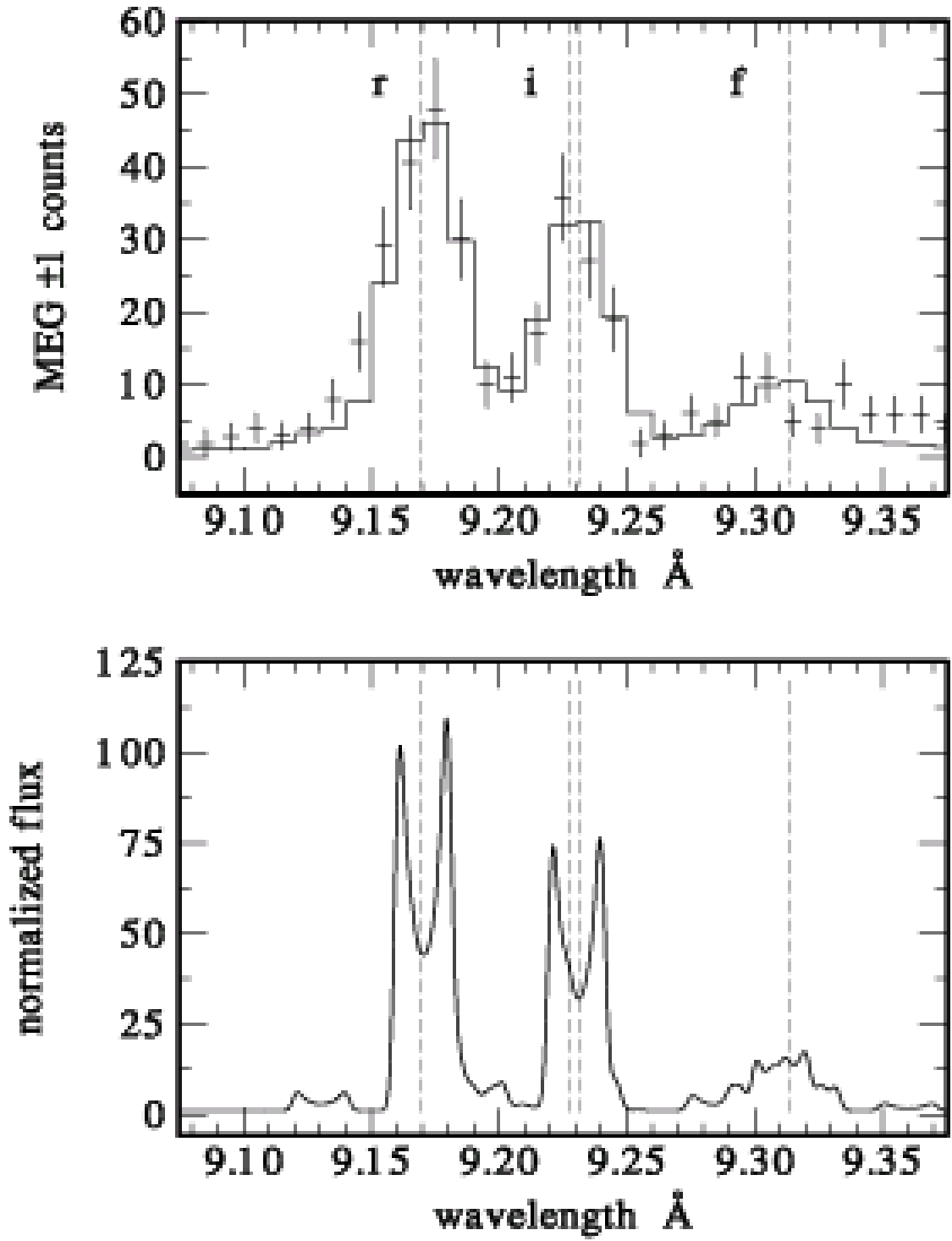}
\caption{
The top panel MTDGD model best fit to the $Chandra$ HETGS MEG +/-1 first order
count spectrum of $\zeta$ Oph in the region of the He-like Mg XI $fir$ lines. The resultant chi-square to the number of degrees of freedom (DOF), i.e., $\chi^2$/DOF , is equal to 33.1/30. The bottom panel shows the
predicted MTDGD model normalized observed flux that was used to generate
the model count spectrum (top panel).  The characteristic double peaked
lines expected from a disk-type structure are clearly evident in the $r$- and $i$-lines.  The $f$- line shows the effect of several weaker satellite lines.
The vertical lines show the rest wavelengths of the $r$-, $i$-, and $f$-lines.
\label{f6}}
\end{figure}

The model Mg XI $fir$ emission lines are calculated by a simple $\phi$ integration
since the radial position of the emission zone is fixed.  The
emission is attenuated by the radial and $\phi$ dependent line-of-sight cool
stellar wind X-ray continuum optical depth through the disk, including the
effects of stellar occultation.  The model X-ray emission includes all
emissivities in a given wavelength region, e.g., the $fir$ lines, their
satellite lines, any other lines that may be present, and the continuum.
Once we specify the location and temperature of the disk distributed X-ray
plasma (as determined by the MTDGD model), along with the effective EUV
density as determined by a known photospheric flux, there is basically only
one single free parameter, 
the normalization factor obtained from the fit
which gives the total emission measure of the integrated disk distributed X-ray sources (i.e., a measure of $n_e^2$ times the volume of the X-ray emitting
plasma).  The predicted HETGS MEG+/-1 counts are compared to the observed
counts in the top panel of Fig.~\ref {f6} (using a bin size of 0.01 $\rm {\AA}$).  The bottom panel of Fig.~\ref {f6} shows the predicted input normalized model flux
used to generate the model MEG+/-1 first order counts by using the extracted
Ancillary Response File (ARF) and Redistribution Matrix File (RMF)
appropriated for the $\zeta$ Oph data. The most obvious feature seen in the $r$-
and $i$-lines is the characteristic double peaked line as expected from
viewing a disk collection of sources seen at a large inclination angle. 
Since there is no radial velocity component in our model, the blue and red
peaks should have the same strength.  However, we see that the red peak of
the $r$-line is slightly larger than the blue peak, which we attribute to the
presence of several weaker satellite lines red-ward of the $r$-line.  The most
notable effect of other lines is seen in the vicinity of the $f$-line where we
see that the double-peaked characteristic in the $f$-line is masked by these
other lines. We also see that there is a slight emission excess on the
blue-side of the $r$-line profile which suggests that the disk confined X-ray
sources may have an outward radial velocity component of a few hundred $\kmsec$,
or perhaps the wind region above the disk might have standard radiation
driven wind shocks that are contributing to the overall emission.
Nevertheless, we conclude that the MTDGD model can reproduce the observed Mg
XI $fir$ line shapes and line strengths, and more importantly, the model
predicts an $f/i$ ratio that is very good agreement with the observations.
Although this model fit to the HETGS data predicts a log EM of 54.52 which
is approximately 40\% larger than the observed emission measure derived from $ROSAT$ observations as listed in Table~\ref {tab2}. An explanation
for this small difference is that $\zeta$ Oph is a variable X-ray source
(Waldron 2005) and at the time of the $Chandra$ observation the X-ray flux is
about 40\% larger than when the star was observed with $ROSAT$.

\section{Discussion and Conclusions}

The magnetically torqued disk model including gravity darkening effects
or MTDGD, has here been tested to see if it can explain the basic
X-ray properties of Oe/Be stars for which the model was developed.
We had already found in the original
papers on the MTD concept by Cassinelli \etal (2002) and Brown \etal (2004)
that the idea of mass in Be disks is channeled by magnetic
fields is consistent with the \Halpha\ luminosities of Be stars and
that the mass in the disks as derived from polarization observations
is also explainable, though we again note that inclusion of
$g_z$ will likely increase these in the model.

We have used a model that explains how matter can enter a disk with
sufficient angular momentum to explain the quasi-Keplerian disks of Be
stars, and the model uses field strengths that are comparable to those being
found for other B stars.  However an essential property of the MTDGD
model is that it requires that X-rays be produced owing to the abrupt
braking of the wind at the shock fronts. In summary, the paper contains several interesting results. (a) The paper tests the prediction that X-rays should be produced by the impact of channeled winds onto a disk. These X-rays would probably not be predicted from other current Be star models such as those in which the disk is produced by an extraction of angular momentum from the surface of a critically rotating star. (b) The model was based on the assumption that Be stars are rotating at their traditional values of about 70 percent critical and these rotation rates were found to be sufficient to explain the X-ray emission within the context of the MTD picture. (c) The model was found to require fields of order $10^2$ Gauss for Be stars, and our results show that these are adequate for the broad band X-ray production. (d) Broad band $ROSAT$ X-ray fluxes can be produced from B1 to B8 with MTDGD model parameters. (e) Fits are achievable even for the late B8 stars without invoking the presence of a dwarf M companion. (f) The model predicted that the Helium-like ion Mg XI had an $\rm R_{fir}$ about 1.8 stellar radii, which agrees well with where that line emission is expected to arise in $\zeta$ Oph with $Chandra$ observations, i.e., the $fir$ lines are predicted to occur there and to be broad, with a half width of about 400 $\kmsec$ and the temperature structure is consistent with the formation of the Mg XI line.

Our discussion thus far has dealt with understanding the fundamental
properties of Be stars as revealed by the $ROSAT$ and $Chandra$ observations. We have raised several 
questions during this paper that can now be addressed. (1) For our latest star B7 IVe, we found that 
the observed $ROSAT$ level of X-rays could be produced if the velocity law had value $\beta =0.77$. 
This small value for $\beta$ means that the channeled wind is colliding with the disk at a larger 
fraction of terminal wind speed than is the case for our other stars, which have $\beta$ values ranging 
from 1.3 to 2.8. The other required parameters for this star seem plausible: $B \sim 50$ Gauss, $S_o \sim 0.6$, 
(in the first of our two fitting procedures). Cohen \etal (1997) suggested that the emission measure 
needed to explain X-rays from late B stars were excessive, and that perhaps X-rays from magnetically 
confined region at the base of the wind are needed. So from our model of this star it appears that a 
magnetically confined X-ray formation region at the base is not needed. A bipolar magnetic field could 
instead just be torquing and channeling the wind toward the disk via X-ray emitting shocks. (2) The sharp 
drop-off of the X-ray luminosity beyond about B2 V, also seems to be explainable with a plausible range of 
our $\beta$, $S_o$, and magnetic field values of about 685 to 130 Gauss for our B2 and B3 stars.

It appears that the MTDGD model has the ability to answer two of the more
difficult questions concerning existing X-ray observations of Be stars, with
plausible parameters. The model can explain the X-ray luminosity across the B spectral band and it can 
explain reasonably well the observed line profile results from $Chandra$. Finally, it is important to 
note that our model results are not dependent
on the nature of the high density regions on the equatorial plane
for which there is controversy regarding field wrapping and magnetic
breakouts. This is because our models are in effect
providing information only about the X-ray formation
regions at the boundaries of the disk, and the flow through these
boundaries occurs before the matter reaches the cooled 
compressed region near the equatorial plane. We are not requiring that
the gas be controlled by a strong field all the way to the equatorial
plane, and in fact think that the gas is no longer dominated by the field
in that region and is free to acquire a quasi-Keplerian orbital motion. 
The needed angular momentum had been transferred to the gas in the pre-shock
magnetically torqued and channeled MTD flow.

\acknowledgements

We would like to thank the anonymous referee for useful comments which led to a significant improvement in the paper. We would like to thank M. Maheswaran for helpful comments and K. Dellenbusch for assistance 
with the initial programming involved. QL was supported in part by NSFC grant-10273002, 10573022, 10778601 
and by the Royal Society Sino-British Fellowship Trust Award.  JPC, QL, NAM and JCB were supported 
in part by the NASA $Chandra$ theory and modeling grant TM3-4001. NAM acknowledges support from a 
Research Corporation award. JCB received support from a UK PPARC research grant. WLW acknowledges 
support by award GO2-3027A issued by the $Chandra$ X-ray Observatory Center. $Chandra$ is operated by 
the Smithsonian Astrophysical Observatory under NASA contract NAS8-03060.


\clearpage

\begin{table}[tbp]
\begin{center}
\caption{The Quantities of Program Stars     \label{tab1}}
\vspace{1ex}
\begin{tabular}{llccccccccr}
\hline\hline
\\
Star & Spectral& log$L_B$ & $T_{e}$ & $R_{\ast}$ &
$M_{\ast}$ & $\dot M$ & $v_{\infty}$ & Distance & $v \sin i$ & \\
Name & Type & ($L_{\odot}$)& (K) & ($R_{\odot}$) &
($M_{\odot}$) & ($M_{\odot}$ yr$^{-1}$) & (km s$^{-1}$) & (pc)& (km s$^{-1}$) & \\
\hline

$\zeta$ Oph & O9.5 Vn& 5.04&  31600& 8.00& 25.0& $4.0\times 10^{-8}$&  1500&   154& 385 & \\

$\kappa$ CMa& B1.5 IVe& 4.21&  24690& 6.84& 12.9& $4.2\times 10^{-9}$&  1560&   308& 200& \\

$\eta$ Cen  &B1.5 Ve& 3.99&  24690& 5.31& 11.2& $1.5\times 10^{-9}$&  1660&   110& 345& \\

$\delta$ Cen& B2 IVe&  4.01&  23010& 6.31&  9.8& $2.0\times 10^{-9}$&  1310&   138& 155& \\

$\mu$ Cen   & B2 IV-Ve& 3.89&  23010& 5.50& 10.4& $8.5\times 10^{-10}$& 1470& 163& 180&\\

$\alpha$ Ara& B2 Ve&  3.77&  23010& 4.79&  9.8& $4.5\times 10^{-10}$& 1540& 122& 315&\\

$\alpha$ Eri& B3 Ve&  3.33&  19320& 4.07&  6.9& $4.2\times 10^{-11}$& 1330&  27& 250&\\

$\alpha$ Col& B7 IVe& 2.45&  12790& 3.39&  3.7& $3.0\times 10^{-12}$& 1250& 44& 210& \\

\hline\hline

\end{tabular}

\end{center}

\tablecomments{
All quantities are taken from \citet{cohen} except $\zeta$ Oph which is from \citet{bergh} 
and Howarth \& Prinja (1989).
}

\end{table}

\begin{table}[tbp]


\begin{center}

\caption{The Derived Quantities of Program Stars for a fixed value of $S_o$ and $\gamma$ with $\beta$ 
varied to fit the log $L_x/L_B$ ratio    \label{tab2}}


\vspace{1ex}

\begin{tabular}{lcccccc|ccc|ccr}

\hline\hline

 & & & & & & &  \multicolumn{3}{c}{\textit{Theoretical}} & \multicolumn{2}{|c}{\textit{Observational}}&

 \\

Star & & &B& & $X_i$ & $X_o$ &log&log & &log&log& \\

Name &$\beta$ &$S_o$& (Gauss) &$\gamma$ &($R_{\ast}$)&($R_{\ast}$)&EM$_t$&$L_{xt}/L_B$&HR$_t$&EM$_o$&$L_{xo}/L_B$&\\

\hline

$\zeta$ Oph &1.29& 0.64& 2072& 1.60&  1.35& 1.86& 53.91& -7.47& 0.630& 54.30& -7.47&\\

$\kappa$ CMa&2.43& 0.42& 917&  2.09&  1.78& 2.47& 52.65& -7.94& 0.521& 52.60& -7.94&\\

$\eta$ Cen  &1.53& 0.69& 514&  1.39&  1.28& 1.81& 51.99& -8.43& 0.397& 51.90& -8.42&\\

$\delta$ Cen&2.81& 0.36& 684&  2.42&  1.98& 2.77& 51.90& -8.57& 0.264& 51.77& -8.57&\\

$\mu$ Cen   &1.34& 0.38& 685&  2.77&  1.91& 2.55& 53.07& -7.01& 0.913& 52.83& -7.01&\\

$\alpha$ Ara&1.61& 0.64& 326&  1.48&  1.35& 1.88& 51.42& -8.78& 0.358& 51.33& -8.78&\\

$\alpha$ Eri&1.45& 0.56& 130&  1.78&  1.47& 2.01& 50.80& -8.89& 0.558& 50.77& -8.89&\\

$\alpha$ Col&0.77& 0.59& 44&   1.95&  1.42& 1.89& 50.41& -8.28& 0.838& 50.50& -8.28&\\

\hline\hline

\end{tabular}

\end{center}

\tablecomments{
We use the average rotation rate and threshold magnetic field as the fixed 
parameters, then fit X-rays by adjusting $\beta$ through the simulations. In the 
table, $X_i$ and $X_o$ represent the inner and outer extent of the disk with stellar 
radius as the unit, respectively. The subscripts `t' and `o' in $EM$, $L_x/L_B$ and HR 
mean the theoretical and observational, respectively.
}

\end{table}

\begin{table}[tbp]


\begin{center}

\caption{The Derived Quantities of Program Stars for a fixed value of $\beta$ and $\gamma$ with $S_o$ 
varied to fit the log $L_x/L_B$ ratio   \label{tab3}}


\vspace{1ex}

\begin{tabular}{lcccccc|ccc|ccr}

\hline\hline

 & & & & & & &  \multicolumn{3}{c}{\textit{Theoretical}} & \multicolumn{2}{|c}{\textit{Observational}}&

 \\

Star & &&B& & $X_i$ & $X_o$ &log&log & &log&log&\\

Name &$\beta$ &$S_o$& (Gauss) &$\gamma$ &($R_{\ast}$)&($R_{\ast}$)&EM$_t$&$L_{xt}/L_B$
&HR$_t$&EM$_o$&$L_{xo}/L_B$&\\

\hline

$\zeta$ Oph &1.0&0.752& 1922& 1.49& 1.21& 1.72& 53.91& -7.47& 0.636& 54.30& -7.47& \\

$\kappa$ CMa&1.0&0.821& 604&  1.38& 1.14& 1.68& 52.65& -7.94& 0.526& 52.60& -7.94&\\

$\eta$ Cen  &1.0&0.879& 477&  1.30& 1.09& 1.66& 51.98& -8.42& 0.440& 51.90& -8.42&\\

$\delta$ Cen&1.0&0.815& 392&  1.38& 1.15& 1.68& 51.90& -8.57& 0.250& 51.77& -8.57&\\

$\mu$ Cen   &1.0&0.495& 546&  2.21& 1.60& 2.11& 53.08& -7.01& 0.906& 52.83& -7.01&\\

$\alpha$ Ara&1.0&0.858& 291&  1.32& 1.11& 1.67& 51.41& -8.78& 0.390& 51.33& -8.78&\\

$\alpha$ Eri&1.0&0.725& 112&  1.54& 1.24& 1.74& 50.81& -8.89& 0.561& 50.77& -8.89&\\

$\alpha$ Col&1.0&0.488& 50 &  2.24& 1.61& 2.13& 50.41& -8.28& 0.850& 50.50& -8.28&\\

\hline\hline

\end{tabular}

\end{center}

\tablecomments{
The same as that in Table~\ref{tab2}, but here we use $\beta=1$ and the 
threshold magnetic field as the fixed parameters instead, then fit X-rays by 
adjusting $S_o$ in the simulations.
}

\end{table}

\clearpage

\end{document}